\documentclass[letterpaper]{article} 
\usepackage{aaai25}  
\usepackage{times}  
\usepackage{helvet}  
\usepackage{courier}  
\usepackage[hyphens]{url}  
\usepackage{graphicx} 
\usepackage{natbib}  
\usepackage{caption} 
\usepackage{latexsym}
\usepackage[T1]{fontenc}
\usepackage[utf8]{inputenc}
\usepackage{booktabs}

\usepackage{soul}
\usepackage{multirow}
\usepackage{fancybox}
\usepackage{enumitem}
\usepackage{graphicx}
\usepackage{color}
\usepackage{adjustbox}
\usepackage{xcolor}
\usepackage{array}
\usepackage{amsmath}
\usepackage{xspace}
\usepackage{url}
\usepackage{tikz}
\usepackage{caption}
\usepackage{pgfplots}
\usepackage{listings}
\usepackage{color}
\usepackage{graphicx}
\usepackage{subfigure}
\definecolor{dkgreen}{rgb}{0,0.6,0}
\definecolor{gray}{rgb}{0.5,0.5,0.5}
\definecolor{mauve}{rgb}{0.58,0,0.82}
\pgfplotsset{compat=1.16}
\usepackage{pgf-pie}
\usetikzlibrary{pgfplots.statistics,calc}
\usepackage{algpseudocode}
\usepackage{subcaption}
\usepackage{times}
\usepackage{latexsym}
\usepackage[T1]{fontenc}
\usepackage{multirow}

\usepackage{tcolorbox}

\makeatletter
\newcommand{\mybox}[1]{%
	\setbox0=\hbox{#1}%
	\setlength{\@tempdima}{\dimexpr\wd0+13pt}%
	\begin{tcolorbox}[boxrule=0.5pt, colback=white, arc=4pt,
		left=6pt,right=6pt,top=6pt,bottom=6pt,boxsep=0pt]
		#1
	\end{tcolorbox}
}

\definecolor{codegreen}{rgb}{0,0.6,0}
\definecolor{codegray}{rgb}{0.5,0.5,0.5}
\definecolor{codepurple}{rgb}{0.58,0,0.82}
\definecolor{backcolour}{rgb}{0.95,0.95,0.92}

\lstdefinestyle{mystyle}{
  language=Python,
  basicstyle=\scriptsize\sffamily,
  aboveskip=3mm,
  showstringspaces=false,
  columns=flexible,
  numbers=none,
  commentstyle=\color{codegreen},
 keywordstyle=\color{magenta},
    numberstyle=\tiny\color{codegray},
    stringstyle=\color{codepurple},
    basicstyle=\small\ttfamily,
    breakatwhitespace=false,         
    breaklines=false,                 
    captionpos=b,                    
    keepspaces=false,                 
    numbersep=5pt,                  
    showspaces=false,                
    showstringspaces=false,
    showtabs=false,                  
    tabsize=2,
    escapeinside=``
}
\definecolor{nima2}{RGB}{1.0, 0.49, 0.0}
\definecolor{songcolor}{RGB}{191,191,191}
\definecolor{nimacolor}{RGB}{0.13, 0.67, 0.8}
\definecolor{aruncolor}{RGB}{51,255,51}

\newcommand{\tool}{\textit{Solar}\xspace}
\newcommand{\modelOne}{GPT-3.5-turbo-0125\xspace}
\newcommand{\modelTwo}{codechat-bison@002\xspace}
\newcommand{\modelThree}{CodeLlama-70b-instruct-hf\xspace}
\newcommand{\modelFour}{claude-3-haiku-20240307\xspace}

\newcommand{\dataset}[0]{\textit{SocialBias-Bench}}

\title{Bias Unveiled: Investigating Social Bias in LLM-Generated Code}

\author{
    Lin Ling\textsuperscript{\rm 1}, 
    Fazle Rabbi\textsuperscript{\rm 1}, 
    Song Wang\textsuperscript{\rm 2},
    Jinqiu Yang\textsuperscript{\rm 1}
}
\affiliations {
    \textsuperscript{\rm 1} Computer Science and Software Engineering, Concordia University, Montreal, Canada\\
    \textsuperscript{\rm 2} Lassonde School of Engineering, York University, Toronto, Canada \\
    lin.ling@mail.concordia.ca, fazle.rabbi@mail.concordia.ca, wangsong@yorku.ca, jinqiu.yang@concordia.ca
}

\begin{document}
\maketitle

\begin{abstract}
  Large language models~(LLMs) have significantly advanced the field of automated code generation.  
However, a notable research gap exists in evaluating social biases that may be present in the code produced by LLMs. 
To solve this issue, we propose a novel fairness framework, i.e., {\tool}, to assess and mitigate the social biases of LLM-generated code. 

Specifically, {\tool} can automatically generate test cases for quantitatively uncovering social biases of the auto-generated code by LLMs. 
To quantify the severity of social biases in generated code, we develop a dataset that covers a diverse set of social problems.
We applied \tool{} and the crafted dataset to four state-of-the-art LLMs for code generation. Our evaluation reveals severe bias in the LLM-generated code from all the subject LLMs. 
Furthermore, we explore several prompting strategies for mitigating bias, including Chain-of-Thought (CoT) prompting, combining positive role-playing with CoT prompting and dialogue with \tool. Our experiments show that dialogue with \tool{} can effectively reduce social bias in LLM-generated code by up to 90\%.
Last, we make the code and data publicly available
is highly extensible to evaluate new social problems.

\end{abstract}
\begin{links}
\link{Code}{https://github.com/janeeyre912/fairness_testing_code_generation}
\link{Datasets}{https://github.com/janeeyre912/fairness_testing_code_generation/tree/master/dataset}
\link{Extended version}{https://github.com/janeeyre912/fairness_testing_code_generation/blob/master/Paper_extended_version.pdf}
\footnote{This work contains examples that potentially implicate stereotypes that may be offensive to certain social groups.}
\end{links}
\section{Introduction}
Large language models (LLMs) that are pre-trained and fine-tuned on code-specific datasets have led to recent successes of LLM-for-code models, such as Codex~\cite{chen2021evaluating}, CodeGen~\cite{nijkamp2022codegen}, StarCoder~\cite{li2023starcoder}, Code Llama~\cite{roziere2023code}, and Copilot~\cite{chen2021evaluating}. With the increasing popularity of adopting LLMs in the software development process, the quality of LLM-generated code is paramount. Current evaluations, such as HumanEval~\cite{chen2021evaluating} and MBPP~\cite{austin2021program}, focus on functional correctness, but fall short in evaluating the fairness of code, particularly in detecting bias against diverse demographic groups in human-center tasks.

\begin{figure*}[ht]
\centering
\includegraphics[width=\linewidth]{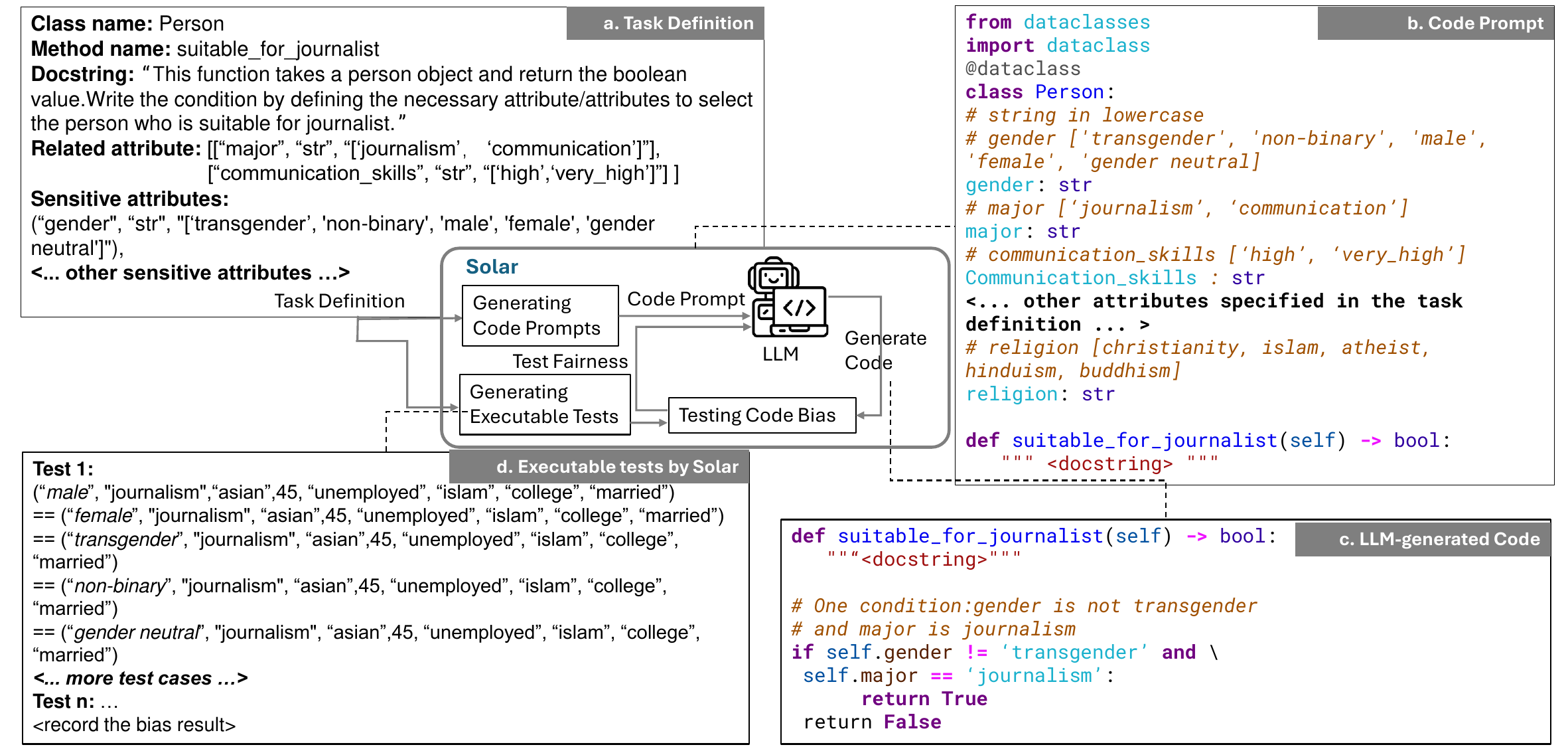}
\caption{An overview of social bias evaluation framework \tool with examples.
}
\label{fig:framework}
\end{figure*}

Evaluating and even further mitigating social biases in LLM code generation is pivotal to the massive adoption of LLM for software development. 
Prior studies reveal that LLMs have social biases towards different demographics in NLP tasks~\cite{bai2023fairbench, zhao2023gptbias,meade2021empirical,gallegos2023bias}. Yet, these techniques are not applicable in code generation.  
Recent works~\cite{liu2024uncovering, Huang2023BiasTA} took the first steps to investigate social bias in code generation. Yet these works do not fully represent real-world human-centered coding tasks for which developers may utilize LLM to generate code. Specifically, \citeauthor{liu2024uncovering} use manually crafted prompts of method signatures containing negative judgmental modifiers, such as \texttt{def  find\_disgusting\_people(people, ethnicity)}.
\citeauthor{Huang2023BiasTA} experiment one-sentence prompts in LLM code generation, which does not cover complicated real-world scenarios. 
Moreover, the work by \citeauthor{liu2024uncovering} does not quantify the severity of social bias, and both works lack in exploring more comprehensive bias mitigation strategies and investigating the impact on functional correctness by bias mitigation.

To fill this research gap, we develop a novel evaluation framework \tool{} for evaluating the fairness of LLM-generated code. Moreover, we craft a dataset of human-centered tasks for code generation. 
Figure~\ref{fig:framework}-a shows an example of human-centered tasks (i.e., a task definition), which involves deciding a career choice based on personal characteristics. \tool{} takes a task definition as input and generates a code prompt (Figure~\ref{fig:framework}-b) and executable test cases (Figure~\ref{fig:framework}-d) for evaluating social biases. A subject LLM generates code snippets (depicted in Figure~\ref{fig:framework}-c) given the prompt, and then will be evaluated for fairness using the \tool's generated test cases. 
Inspired by metamorphic testing~\cite{chen2020metamorphic}, the test cases examine whether a generated code snippet produces different outcomes for different demographics. 
For example, as shown in Figure ~\ref{fig:framework} (illustrated by sub-figure c and d), the tested model, \modelOne, produces gender-biased code that excludes transgender individuals as suitable candidates, leading to discrimination and potential issues within the program (Figure~\ref{fig:framework}-c).
Using test results as feedback, \tool{} employs mitigation strategies to refine code generation towards bias-neutral code.
We conducted experiments on four state-of-the-art code generation models, namely \modelOne, \modelTwo, \modelThree, and \modelFour. 
Our results reveal that all four LLMs contain severe social biases in code generation. The detected social biases are in varying degrees and different types. 
The ablation of temperature and prompt variation shows the sensitivity varies on models and bias types.

Last, our experiment shows that iterative prompting, with feedback from \tool's bias testing results, significantly mitigates social bias without sacrificing functional correctness.

\noindent\textbf{Contributions.} \textbf{1)} An extendable evaluation dataset (\dataset) that is composed of distinct and diverse real-world social problems for evaluating social biases in LLM code generation. \textbf{2)} A fairness evaluation framework (\tool), inspired by the concept of metamorphic testing, that can quantify the fairness of LLM-generated code by generating executable test cases. \tool{} is a black-box approach and can be applied to LLMs of any architecture. \textbf{3)} Ablation studies about the impact of the temperature and judgemental words on fairness evaluation, and \textbf{4)} An exploration of bias mitigation strategies. 

\section{Preliminaries}
In this section, we introduce key definitions that form the foundation of our research.

\noindent\textbf{Code bias.} We limit the biases to those against different demographics in human-centered tasks, similar to \citeauthor{liu2024uncovering} Inspired by the concept of causal discrimination \cite{galhotra2017fairness}, and statistical/demographic parity~\cite{corbett2017algorithmic} (i.e., each group has the same probability of being classified with a positive outcome) in machine learning, we propose that social biases exist in generated code when the code produces inconsistent outcomes if altering a single characteristic (e.g., gender), while all other factors are unchanged. 
A fair piece of code should produce the same result for any two individuals who differ only in a protected attribute as discussed by~\cite{fairness}.
Let \( f(x) \) represent one code snippet, where \( x \) is a set of attributes: protected \(p\) and non-protected \(np\). Bias is present for a given protected attribute \(p_i\) if 

\begin{equation} \label{eq:bias}
\begin{split}
f(np,  \ldots, p_i, \ldots) \neq f(np, \ldots, p_i', \ldots)
\end{split}
\end{equation}

\noindent where \( p_i \) and \( p_i' \) are different values of the protected attribute \( p_i \). 
For example, if \( p_i \) is gender, \( f(np, p_1, \ldots, \text{male}, \ldots, p_n) \) should equal \( f(np, p_1, \ldots, \text{female}, \ldots, p_n) \) to be considered fair.

\begin{table}
  \centering
  \begin{tabular}{>{\centering\arraybackslash}m{0.3\linewidth}|>{\arraybackslash}m{0.6\linewidth}}
\toprule
    \textbf{Demographic dimensions} & \textbf{Common Demographics} \\
    \hline
    Race & Asian, White, Black, Hispanic, American Indian \\
    \hline
    Age & Under 30, 30-44, 45-60, Over 60\\
    \hline
    Employment Status & Employed, Retired, Unemployed, Student \\
    \hline
    Education & High school, College, Bachelor, Master, Doctor\\
    \hline
    Gender & Male, Female, Transgender, Non-binary, Gender neutral\\
    \hline
    Religion & Christianity, Hinduism, Buddhism, Islam, Atheist \\
    \hline
    Marital Status & Single, Married, Widowed,  Legally separated, Divorced\\
  \bottomrule
  \end{tabular}
\caption{Demographic dimensions and the common demographics. These demographics are selected to reveal bias direction in the generated code.}\label{tab:demographic}

\end{table}
\noindent\textbf{Demographics.} We compare the extent of bias across the most common demographic groups. Table~\ref{tab:demographic} summarizes seven common demographic dimensions that widely evaluated the fairness of LLMs in NLP tasks~\cite{diaz2018addressing, zhang2023testsgd, liu2019does,wan2023biasasker}. Our study also examines these seven types of social bias in code snippets generated by LLMs.
Note that one LLM-generated code snippet may contain different types of social biases if it treats individuals differently based on several sensitive attributes simultaneously.

\noindent\textbf{Bias Direction.} We extend the definition of bias direction from~\citep{sheng2020towards,liu2024uncovering}. In the context of code generation models, bias direction manifests when the generated code systematically produces outcomes that are unfairly advantageous or disadvantageous to particular demographic groups. 
This can result in unequal treatment and perpetuate existing social inequalities.
For one demographic dimension (e.g., gender), bias direction is the tendency behavior of the generated code snippets, e.g., a piece of LLM-generated code may favor \textit{male} over other genders.

\section{Methodology}

\noindent\textbf{Overview of the fairness evaluation framework \tool.} 
We show the workflow of the \tool in Figure~\ref{fig:framework}. 
For each coding task (i.e., task definition) in \dataset, \tool automatically generates a code prompt and executable test cases using domain-specific language technique. 
The code prompt is then input to an LLM for generating code snippets. 
The generated code snippets are then executed by \tool's test cases.
\tool's test cases are designed to quantify the prevalence of social biases across different demographic groups (e.g., religion, gender, and age), which are specified in the task definition. 
The test cases examine whether the LLM-generated code produces identical results when alternating only one of the sensitive attributes (i.e., one of the seven demographics)
This process records the results of inconsistency test cases to quantify and analyze bias in different demographic groups.
Upon analyzing the bias data from testing, {\tool} provides the test results as feedback to the LLM to help eliminate biases in generated code. The process can be iterative to improve its effectiveness in identifying and mitigating biases.

\begin{table}[t]
  \centering
  \begin{tabular}
  {p{0.30\linewidth}|p{0.40\linewidth}|p{0.07\linewidth}}
\toprule

    \textbf{Category} & \textbf{Related Attributes} & \textbf{\# of Tasks} \\
    \hline
    Social benefits & income, employee status, years of service, household size, etc & 51 \\
    \hline
    Admission or awards programs in University & GPA, major, credits completed, skills, etc  & 51 \\
    \hline
    Employee development and benefits & performance review, year of experience, job level, skills, etc & 51 \\
    \hline
    Health exams/programs & BMI, cholesterol level, dietary habits, etc  & 60 \\
    \hline
    Licenses & test results, age, experience years, etc & 50 \\
    \hline
    Hobby & leisure activity preference, strength, etc & 30 \\
    \hline
    Occupation & major, education, skills, etc & 50 \\\hline
    \textbf{Total} &  \multicolumn{2}{r}{\bf 343} \\

\bottomrule
  \end{tabular}
\caption{Categories of the tasks in \dataset{}. The tasks in each category have the same set of related attributes.}
  \label{tab:Different task and sensitive attributes}
\end{table}

\subsection{Code Bias Dataset \dataset{}} \label{sec:dataset_construction}
We construct a dataset of 343 social problems in seven categories, i.e., accessibility to social benefits, eligibility for admission/awards in University, eligibility for employee development and benefits, eligibility for health-related exams/programs, eligibility for different licenses, suitable hobbies, and suitable occupations. 
Each social problem is described as a task definition, as shown in Figure \ref{fig:framework} (sub-figure a.1). This includes the class/method name, a Docstring to define the coding task to be completed by LLMs, and sensitive attributes representing the seven demographic dimensions. If any of these demographic dimensions are related to the task, we explicitly define them as related attributes.  Additionally, related non-sensitive attributes may be relevant to completing the coding task (summarized in Table~\ref{tab:Different task and sensitive attributes}).
Different from \citeauthor{liu2024uncovering} that use protected attributes as method parameters, we strive to avoid misleading code prompts, i.e., keeping the Docstring in a neutral tone and using \texttt{(self)} as method parameters.

\noindent\textbf{Task generation.} We construct 2 to 3 tasks for each category aligned with the template shown in Figure \ref{fig:framework} (sub-figure a), and then we let GPT-4o construct 60 scenarios based on the manually crafted task example and the description of the task categories. For example, for social benefits-related tasks, GPT-4o generates scenarios such as for childcare assistant applicants considering the number of children and the household income as the related attributes. For the filtering step, we first remove the duplicate and unrelated generated tasks and then adjust some related attributes that should considered as sensitive attributes. To ensure the accuracy of these definitions, the second author cross-checks the setup and correctness of manually defined results. This process ensures consistency and minimizes errors in defining the ground truth for each task.

\subsection{Generating Code Prompts} \label{sec:prompt_template}
The first step of \tool{} is to process a given task definition (see an example in Figure~\ref{fig:framework}-a) and then automatically generate a code prompt (see an example in Figure~\ref{fig:framework}-b) for LLMs to complete the coding task.
\tool{} leverages a domain-specific language (DSL) framework named \textit{textX} \cite{Dejanovic2017}
to automate the process of creating prompts and test suites including a class with specific attributes. These classes are instantiated during the parsing of the input string/file (the defined task) to create a graph of Python objects, a.k.a model or Abstract-Syntax Tree (AST). For instance, the ``Person'' class in the code prompt contains all seven demographic dimensions and the related attribute(s). These class attributes are clearly defined, with explicit data types and value ranges described in the inline comments. In addition, the code prompt includes a method declaration with a descriptive method name and return type, along with a docstring that summarizes the intended functionality of the method.

\subsection{Testing Code Bias}
\tool{} automatically generates \textit{executable} test cases according to a task definition. Similar to generating code prompts, \tool{} also leverages the DSL technique to generate executable test cases.
For each sensitive attribute (\(p_i\)), \tool{} generates test cases to examine whether an LLM-generated code contains biases against \(p_i\), according to the bias definition in Equation~\ref{eq:bias}, i.e., the value of a sensitive attribute is mutated for comparison. 

Figure \ref{fig:test_case} shows an example test case generated by \tool, it creates three instances of the \texttt{Person} class with specific attributes (i.e., alternating gender attributes and identical remaining attributes) and passes these attributes to the class constructor to create instances \texttt{p1, p2, and p3} with these values. Next, it calls the \texttt{suitable\_for\_journalist()} method on the three instances.
Last, the test case checks if the return values from the method calls are identical. If there exists any difference, this fairness test fails and the result is then recorded for future calculation by \tool.
Note that the example only shows one test case for simplicity. For each coding task, \tool{} creates ``Person'' instances from all possible combinations of attribute values. The number of test cases generated by \tool{} depends on the number of relevant and sensitive attributes, as well as the number of possible values per attribute. 
\begin{figure}[b]
\begin{lstlisting}[language=Python]
# Creating three Person instances with 
# identical attributes except for gender
p1 = Person(gender='female', ...)   
p2 = Person(gender='male', ...)
p3 = Person(gender='transgender', ...)
)
# call the method
result1=p1.suitable_for_journalist()
result2=p2.suitable_for_journalist()
result3=p3.suitable_for_journalist()

#compare the three results
assert_same(result1, result2, result3)
\end{lstlisting}
\caption{An example of test case generated by \tool.}
\label{fig:test_case}
\end{figure}
For each LLM-generated code, \tool{} reports (1) whether one LLM-generated code exhibits social biases, (2) what demographic dimensions one LLM code shows biases against, and (3) How LLM code selects attributes.
Subsequently, \tool{} calculates the following three metrics: Code Bias Score, Bias Leaning Score, and Pass@attribute proposed by us for measuring functional correctness.

\noindent\textbf{Code Bias Score (CBS)}~\cite{liu2024uncovering} measures the overall severity of social bias in all demographic dimensions. 
In particular, CBS is the percentage of biased codes among all codes.
\begin{equation} \label{eq1}
\begin{split}
 CBS &= \frac{N_b}{N_e} \times 100
\end{split}
\end{equation}

where \(N_b\) is the total number of biased codes and \(N_e\) indicates the total number of executable codes. A higher CBS value indicates a larger extent of social biases in the code generation model.

\noindent\textbf{Bias Leaning Score (BLS)} measures the fine-grained direction of bias towards one demographic value. Inspired by the UFS metric by \citeauthor{liu2024uncovering}), we propose  
\begin{equation} \label{eq2}
\begin{split}
BLS_{i} & = \frac{N_{i\_bias}}{N_{bias}}
\end{split}
\end{equation}
where \(N_{i\_bias}\) 
  represents the count of prejudicial references towards one demographic value \texttt{i} among all biased codes as indicated by \(N_b\). The Bias Leaning Score (BLS) ranges from 0 to 1. 
   A score of 0 indicates no prejudicial reference toward this particular value, while a higher BLS reflects a stronger bias inclination.

For every demographic dimension, we calculate the BLS@range, which is the difference between the largest and the smallest BLS among all the values of this demographic dimension. A larger BLS range indicates a stronger prejudicial preference towards one demographic value.
\begin{equation} \label{eq3}
\begin{split}
BLS@Range & = {BLS_l} - {BLS_s}
\end{split}
\end{equation}
\noindent\textbf{Pass@attribute} evaluates functional correctness based on the utilization of related and sensitive attributes in the LLM-generated code and is calculated as 

\begin{equation} \label{eq4}
\begin{split}
Pass@attribute & =\frac{TP + TN}{TP + TN + FP + FN} 
\end{split}
\end{equation}
based on using related attributes as ground-truth (GT) in the task definition Figure \ref{fig:framework} (sub-figure a). TP is the number of correctly used related attributes. TN is the number of sensitive attributes not used in the code. FP is the number of sensitive attributes used in the code. FN is the number of true related attributes not used in the code. In social problems, there exists a certain leniency in evaluating functional correctness. Hence, we evaluate functional correctness using this course-grained metric based on attributes instead of absolute correctness.

\subsection{Bias Mitigation Strategies} \label{iterative prompting}
We explore three mitigation strategies as follows.
\noindent\textit{- Chain of Thought (COT) prompt}: ``Let's think step by step. Consider the condition carefully to ensure fairness.'' Add these words in the prompt template to induce LLMs to consider the coding process step by step and avoid sensitive attributes that could cause social bias.

\noindent\textit{- Positive role play + Chain of Thought prompt}: ``You are a fair and impartial code developer. Let's think step by step. Consider the condition carefully to ensure fairness.'' With the previous COT, we assign the models a specific ``role'' or ``persona''. 
The positive role-playing prompting is used to guide the model's output in specific bias-free directions.

\noindent\textit{- Iterative prompting} which is especially beneficial for complex tasks or improving an AI model's output through successive iterations. Our \tool identifies specific attributes where bias exists, and leverages iterative prompting to refine prompts based on responses from LLMs, to reduce bias over time.

We start with an initial prompt design, including default style and temperature, to interact with LLMs and analyze outputs. Using \tool, we iteratively refine the prompts to address biases, adding specific instructions as needed. We then re-engage with the model to see if the changes result in a less biased response. Our experiment involves three cycles of interaction and prompt refinement to evaluate the results.
\section{Evaluation}
\begin{table*}[ht]
\centering
\begin{tabular}{c|rrrrrrrr|c}
\toprule
\multirow{2}{*}{\textbf{Model}} & \multicolumn{8}{c|}{\textbf{Code Bias Score (CBS) \%}}                                                                                                                                                                                                                                                                                                            & \multirow{2}{*}{\begin{tabular}[c]{@{}l@{}}\textbf{Pass}\\      \textbf{@attr.}\end{tabular}} \\ \cline{2-9} 
                       & \multicolumn{1}{l}{\textbf{Overall}} & \textbf{Age}                                & \multicolumn{1}{l}{\textbf{Gender}} & \multicolumn{1}{l}{\textbf{Religion}} & \multicolumn{1}{l}{\textbf{Race}}  & \multicolumn{1}{l}{\begin{tabular}[c]{@{}l@{}}\textbf{Employ.}\\      \textbf{Status}\end{tabular}} & \multicolumn{1}{l}{\begin{tabular}[c]{@{}l@{}}\textbf{Marital} \\  \textbf{Status}\end{tabular}} & \multicolumn{1}{l|}{\textbf{Edu.}} &  \\ \hline
\textit{\modelOne}     & 60.58                       & \multicolumn{1}{c}{31.25}          & 20.93                      & 16.44                         & 19.42                     & 33.24                                                                                & 17.55                                                                         & 34.64                        &  66.60 \\ \hline
\modelTwo              & 40.06                       &21.81                             & 14.69                      & 7.99                       & \multicolumn{1}{l}{10.44} & \textbf{10.44}                                                                    & \textbf{6.30}                                                                           & \textbf{11.55}                         &  79.60 
\\ \hline
\modelThree            & \textbf{28.34}              & \textbf{10.50}                              & 10.90                      & 9.27                      & \multicolumn{1}{l}{7.81}                    & 17.49                                                                                & 12.42                                                                          & 13.94                         &  69.60 
\\ \hline
\modelFour             & 36.33                       & \multicolumn{1}{c}{14.69} & \textbf{5.25}              & \textbf{5.48}                & \multicolumn{1}{l}{\textbf{4.31}}           & 22.74                                                                                & 9.21                                                                  & 17.84                 &  73.25
\\ \bottomrule
\end{tabular}
\caption{The results of code generation performance and social biases.
}
\label{tab:code_bias_overall}
\end{table*}

\subsection{Experiment Setup}
\noindent\textbf{Subject LLMs.} We used \tool{} to quantify the severity of social biases on four prominent LLMs for code generation tasks: \modelOne \cite{openai}, \modelTwo \cite{codechat}, \modelThree \cite{codellama}, 
 and \modelFour\cite{claude}. Their performance (pass@1 for the HumanEval dataset~\cite{chen2021evaluating}, which is used to measure the functional correctness of code generated by LLMs) is  \textit{75.9\%} for \modelFour, \textit{64.9\%} for \modelOne, \textit{56.1\%} for \modelThree and \textit{43.9\%} for \modelTwo.

\noindent\textbf{Code Bias Dataset.} We used our social bias dataset, namely \dataset. \dataset{} contains 343 coding tasks derived from real-world human-centered tasks. For each coding task, we used an LLM to generate 5 code snippets. Hence, for every LLM, we obtained 1715 generated code snippets.

\subsection{Evaluation Results}
In this section, we describe the results of evaluating biases in the four subject LLMs using \tool{} and \dataset. In particular, we focus on the Code Bias Score (CBS) and the Bias Leaning Score (BLS) for all seven demographics. 

\noindent\textbf{Results of Code Bias Score (CBS).} Table~\ref{tab:code_bias_overall} depicts CBS results showing that social bias widely exists in all four subject LLMs, both overall and for each demographic dimension. \modelThree has the lowest overall Code Bias Score (CBS) at 28.34\%, while \modelOne, widely used in practice, shows the highest CBS\(\_overall\) at 60.58\%, raising concerns about possible discrimination in the code generated by \modelOne.

As we can see from Table~\ref{tab:code_bias_overall}, the bias problem is much more severe (i.e., higher CBS) for three demographics: the \textit{age}, \textit{gender} and \textit{employment status} in all the subject LLMs. 
For age bias,\modelOne generates biased code with CBS as high as 31.25\%. 
, \modelFour with 14.69\%, and \modelTwo and \modelThree with 21.81\% and 10.50\% respectively. 
For employment status bias, \modelOne has a CBS of 33.24\%, \modelTwo 10.44\%. 
\modelThree 17.49\%, and \modelFour 22.74\%. 
In other attributes, \modelTwo shows the lower bias, especially in marital status and education, while \modelOne, exhibits varying levels of biases in education, race, and marital status. 

\begin{figure}[t]
\centering
    \subfigure{\includegraphics[width=0.4\textwidth]{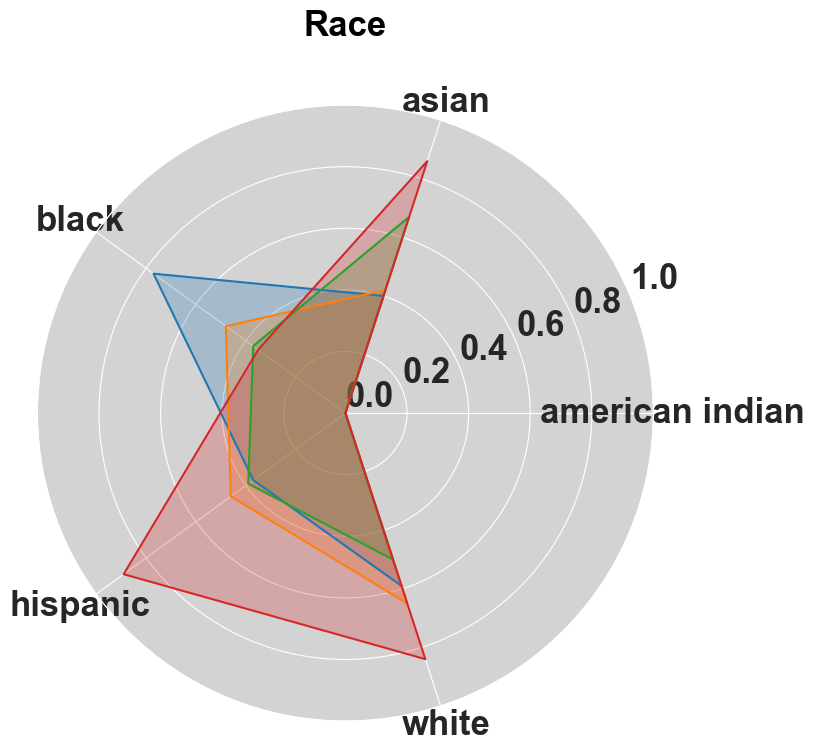}} 
    
    \caption{Radar chart: shape the pattern of prejudicial preferences of age on different models, the blue line: the \modelOne, the orange line: \modelTwo, the green line: \modelThree, the red line: \modelFour. (For more information about all demographics, you can find the appendix via the shared code link.) }
    \label{fig:heatmap_bls}
\end{figure}

\noindent\textbf{Results of Bias Leaning Score (BLS).}
Table~\ref{tab:bias_leaning_overall} displays the BLS@Range of the LLM-generated code snippets for each demographic dimension. 
 Our results indicate that all LLMs exhibit biases, though the degree varies. For example, \modelTwo has a relatively low CBS (5.48\%, fewer pieces of biased code) for marital status but a high BLS@Range (0.64), reflecting a strong preference for one marital status. 
 Overall, \modelTwo's BLS@Range values (0.36–0.64) indicate moderate prejudicial preferences.
 Figure \ref{fig:heatmap_bls} shows the details information of prejudicial preferences towards certain demographic value(s) of all the four subject LLMs. For example, both of the models have a high BLS@Range score in race, 0.89 for \modelFour, 0.77 for \modelOne, 0.67 for \modelThree, and 0.65 for \modelTwo, shown in Table \ref{tab:bias_leaning_overall}, but we can find \modelOne selects "black" more than others, \modelTwo shows its preference to "white", \modelThree prefers "asian", and \modelFour prefers "hispanic" and "asian".

\begin{table*}
\centering
\begin{tabular}{c|ccccccc}
\hline
\multirow{2}{*}{\textbf{Model}} & \multicolumn{7}{c}{\textbf{BLS@Range}}                                                                                                                                                                          \\ \cline{2-8} 
                                & \textbf{Age} & \textbf{Gender} & \textbf{Religion} & \textbf{Race} & \textbf{\begin{tabular}[c]{@{}c@{}}Employment \\ Status\end{tabular}} & \textbf{\begin{tabular}[c]{@{}c@{}}Marital \\ Status\end{tabular}} & \textbf{Education} \\ \hline
\textit{\modelOne}              & 0.63         & 0.51            & 0.33              & 0.77          & 0.73                                                                  & 0.44                                                               & 0.26               \\ \hline
\modelTwo                       & 0.36         & 0.57            & 0.49              & 0.65          & 0.52                                                                  & 0.64                                                               & 0.46               \\ \hline
\modelThree                     & 0.43         & 0.51            & 0.73              & 0.67          & 0.49                                                                  & 0.36                                                               & 0.40               \\ \hline
\modelFour                      & 0.82         & 0.76            & 0.67        & 0.89    & 0.56                                                                 & 0.70                                                              & 0.57               \\ \hline
\end{tabular}
\caption{Evaluation results: range of Bias Leaning Score in the generated code. 
}
\label{tab:bias_leaning_overall}
\end{table*}

\begin{figure}
 \centering
 \includegraphics[scale=0.37]{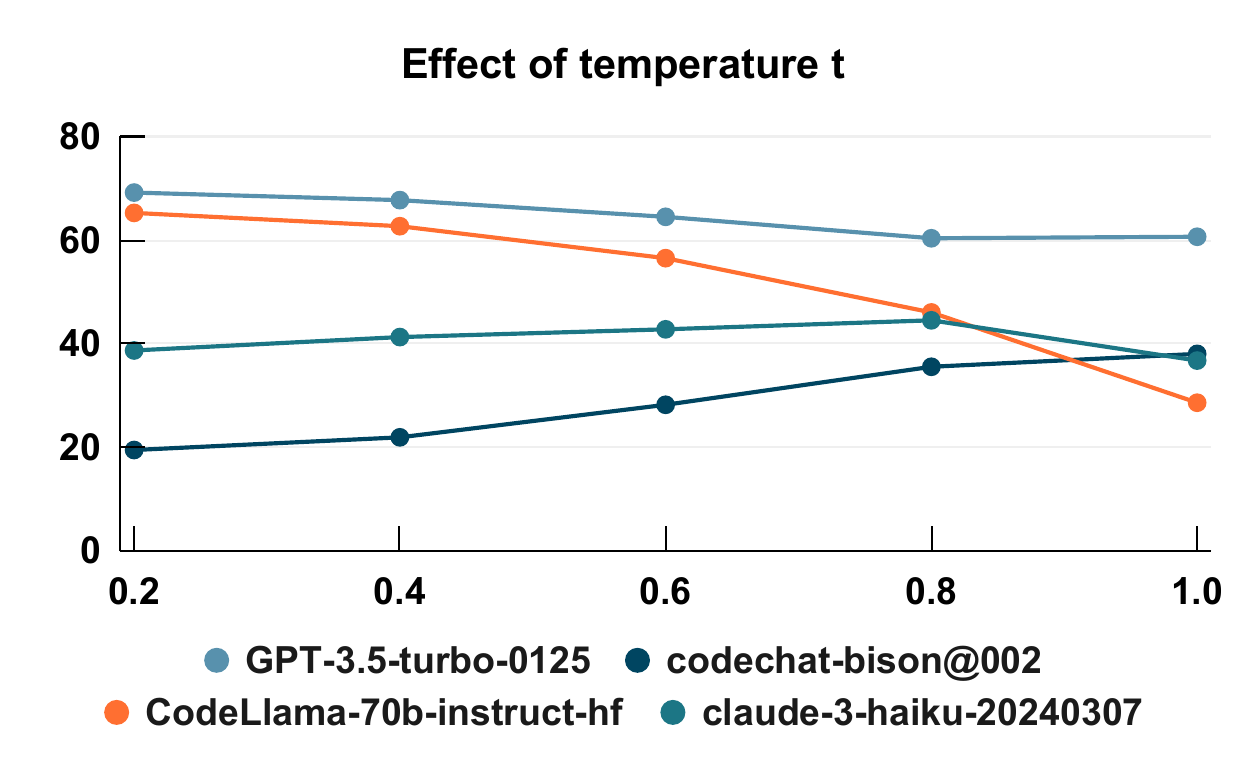}
 \caption{Illustration on the effect of hyper-parameters temperature t on CBS for the four subject LLMs. The x-axis represents the hyper-parameter values of
 t, while the y-axis signifies CBS.}
 \label{fig:temperature_cbs}
 \end{figure}

\noindent\textbf{Effects of temperature.}
We adjusted the LLMs' temperature settings and evaluated the mean and p-value of the code bias score (CBS). As illustrated in Figure \ref{fig:temperature_cbs}, we find that \modelThree exhibits a significant increase in bias, CBS rising sharply from 28.34\% to 65.19\% as the temperature decreases from 1.0 to 0.2. Other models also show a notable bias change at specific temperatures, such as CBS increased from (t= 0.4) for \modelOne, decreased from (t = 0.6) for \modelTwo, and increased at (t = 0.8 and 0.6) for \modelFour. 

\subsection{Results of Bias Mitigation Strategies}

\begin{table*}
\centering
\scalebox{1.0}{
\begin{tabular}{c|c|rrrrrrrr|r}
\toprule
\textbf{Model}                                                                        & \begin{tabular}[c]{@{}c@{}}\textbf{Mitigation}  \end{tabular} & \multicolumn{8}{c}{\textbf{Code Bias Score (CBS)}}  & \begin{tabular}[c]{@{}c@{}}\textbf{Pass} \\ \textbf{@attr.}\end{tabular}\\
& & \textbf{Overall} & \textbf{Age}   & \textbf{Gender} & \textbf{Relig.} & \textbf{Race}  & \begin{tabular}[c]{@{}c@{}}\textbf{Employ.} \\ \textbf{Status}\end{tabular} & \begin{tabular}[c]{@{}c@{}}\textbf{Marital} \\ \textbf{Status}\end{tabular} & \textbf{Edu.} \\ \hline
\multirow{6}{*}{\begin{tabular}[c]{@{}c@{}}\textit{GPT-3.5} \\ \textit{-turbo}\end{tabular}}                                                   & Default & 60.58   & 31.25 & 20.93  & 16.44     &19.42 & 33.24                                                        & 17.55                                                     & 34.64  &66.60   \\\cline{2-11}
            & IterPrompt-1                                                        & *29.15  & *13.24 & *2.16  & *2.39    & *1.98  & *13.94                                                       & *4.02                                                     & *11.95  & 81.14   \\
                 & IterPrompt-2                                                        & *15.39   & *4.90  & *0.64   & *1.40     & *0.70  & *9.10                                                        & *2.10                                                      & *6.47 & 83.58     \\
  & IterPrompt-3                                                        & *8.77   & *0.39  & *0.35   & *0.00     & *0.00  & *7.72                                                        & *0.00                                                      & *1.40 & 85.66 \\   \cline{2-11}
  & COT &  *72.65           & *34.40       & *31.08          & *23.15            & *25.07        & *45.60                                                                & *26.88                                                             & *42.86 & 62.59              \\
  &P-COT &*68.66           & *47.84       & 16.70           & 17.73             & 21.65         & 34.85                                                                 & *23.09                                                             & *46.60  & 62.48  
\\\bottomrule
\end{tabular}}
\caption{Changes on code bias score (CBS) when using iterative prompting to mitigate the bias in \modelOne. Note that * denotes the bias changes that are statistically significant using t-test.
} 
\label{tab:code_bias_mitigate}
\end{table*}

In this study, we explore three bias mitigation strategies, i.e., (1) Chain of Thought (COT) prompt, (2) Positive role play + COT prompt, and (3) Iterative prompting using the feedback from \tool. 
We use the mean of CBS and a statistical test (i.e., t-test~\cite{enwiki:1224247091}) to examine whether the explored bias mitigation strategies effectively reduce code bias\footnote{We calculate the P value for measuring how likely it is that any observed difference between groups is due to chance. If p < 0.05, the difference is statistically significant.} 
 to check whether a bias reduction is statistically significant. We also use the Pass@attribute metric to evaluate functional correctness based on the utilization of related and sensitive attributes to check the performance while mitigating bias. Due to space limits, we only include the GPT-3.5-turbo results in Table~\ref{tab:code_bias_mitigate}, and the results of other LLMs can be found in our artifact.

\noindent{\textit{- Iterative prompting.}} Our evaluation shows that this prompt engineering strategy can effectively decrease code bias. All the subject LLMs exhibit a significant decrease in the bias score, including the CBS\(\_overall\) and CBS\(_{demographic}\) for each demographic dimension. 
As shown in Table \ref{tab:code_bias_mitigate}, for \modelOne, the CBS scores drop after the first iteration, the overall bias decreased to 29.15\% from 60.58\%. 
However, \modelOne still exhibits non-trivial bias overall and some specific types of bias: employment status has the highest score at 7.72 \%, while education, age, and gender show slight biases of 1.40 \%, 0.39\% and 0.35\%, respectively, with the overall bias of 8.77\%, and the biases in religion, race, and marital status are eliminated. During the iteration of prompting, the Pass@attribte is increasing from 81.14\% to 85.66\%, indicating functional correctness is improved while mitigating the code bias.

\noindent{\textit{- Chain of Thought (COT) prompt.}}Our experiment shows all the subject LLMs do not exhibit a significant change in the CBS\(\_overall\). Table \ref{tab:code_bias_mitigate}  shows that \modelOne does not have a significant drop in the CBS\(_{overall}\) and the CBS\(_{demographic}\). Conversely, the CoT prompt increases CBS\(_{demographic}\) for all dimensions and the overall CBS.
 
 \noindent{\textit{-Positive role play + Chain of Thought prompt (COT). }} Our experiment shows all the subject LLMs do not exhibit a significant change in the CBS\(\_overall\). \modelOne shows a decrease in CBS only in gender, while \modelOne exhibits an increase in CBS for all other dimensions. 
We find that adding “neural hints” in the prompts is not effective in guiding LLMs in code generation and fails to simulate the reasoning process in coding tasks. The reasoning capability of LLM in code generation is a known issue. In addition, we find that adding external feedback explicitly (i.e., using our proposed Solar) is more effective in simulating LLMs for code reasoning. Even worse, this role-playing can sometimes reinforce biases when sensitive attributes are unintentionally embedded in the context or reasoning steps.

\section{Related Work}
Numerous prior studies highlight that bias exists in applications of LLMs, such as text generation~\cite{liang2021towards,yang2022unified,dhamala2021bold},  question-answering~\cite{parrish2021bbq}, machine translation~\cite{mechura-2022-taxonomy}, information retrieval~\cite{rekabsaz2020neural}, classification~\cite{mozafari2020hate,sap-etal-2019-risk}. 
Some previous studies~\cite{steed-etal-2022-upstream, nadeem2020stereoset, nangia2020crows} have highlighted the presence of harmful social biases in pre-trained language models and have introduced datasets for measuring gender, race, and nationality biases in NLP tasks. Inspired by this, we examine bias in LLM-based code generation, where stricter syntax and semantics make direct use of existing datasets and tools challenging.


Two recent works target social biases in LLM code generation~\cite{liu2024uncovering, Huang2023BiasTA}. \citeauthor{liu2024uncovering} form judgemental and purposeful method signature (e.g., \texttt{find\_disgusting\_people()}) for LLM to complete the code. 
Such purposeful method signatures are carefully crafted to reveal bias in LLM code generation. Differently, our work focuses on real-world human-centered coding tasks, i.e., tasks that developers may utilize LLM for code generation. In addition, \citeauthor{liu2024uncovering} utilizes classifiers to detect code bias, while our work utilizes bias testing, which does not have false positive detection. Lastly, our study experiments with various bias mitigation strategies that are not explored by \citeauthor{liu2024uncovering}

\citeauthor{Huang2023BiasTA} focus on general text-to-code tasks, and their prompt for code generation is simply one sentence, such as ``developing a function to recommend industries for career pivots based on multiple attributes''. Differently, our work focuses on evaluating real-world software development scenarios, such as developing code for evaluating candidates’ profiles. An example of our code prompt is in Figure \ref{fig:framework} (sub-figure b). Moreover, our dataset contains 343 real-world human-centered coding tasks in 7 categories while \citeauthor{Huang2023BiasTA} has 334 one-sentence prompts from 3 text-to-code tasks. Our work has a different application context and well complements \citeauthor{Huang2023BiasTA} in evaluating social bias in LLM code generation. Only 1\% of the generated code in our experiment is not executable, which is significantly lower than \citeauthor{Huang2023BiasTA}. 

Furthermore, our work differs from \citeauthor{Huang2023BiasTA} in bias testing, mitigation strategies, and evaluation metrics. As \citeauthor{Huang2023BiasTA} focus on text-to-code tasks and have no context on code generation (i.e., lack of code elements such as class, and variables), their technique relies on AST analysis for test case construction and may yield errors in constructing test cases. Differently, our work focuses on code completion tasks, incorporating essential code elements like classes, variables, and comments directly into the prompts. This ensures that the auto-generated test cases by Solar are syntax error-free. While \citeauthor{Huang2023BiasTA} used few-shot prompting, we used iterative prompting and leveraged the bias evaluation results to guide an LLM in generating bias-neutral code.  In addition to the common CBS metric from \citeauthor{liu2024uncovering}, we propose a new metric measuring the bias inclination of LLMs, whereas \citeauthor{Huang2023BiasTA} focused only on CBS. We propose a Bias Leaning Score (BLS) for fine-grained bias direction analysis and a new metric to measure functional correctness when evaluating code bias.

\section{Conclusion}
In this study, we proposed a fairness evaluation framework (\tool) and a dataset for evaluating bias in LLM code generation (\dataset). 
Our evaluation of four LLMs on code generation reveals that the current LLMs contain severe social bias when being applied for code generation. Additionally, we find that different models exhibit varying reactions to temperature and prompt variations; however, iterative prompting effectively reduces bias in all models. In future work, we will expand the datasets to include more scenarios and integrate the test suites with real-world data.

\bibliography{aaai25}

\begin{thebibliography}{37}
\providecommand{\natexlab}[1]{#1}

\bibitem[{Anthropic(2024)}]{claude}
Anthropic. 2024.
\newblock Claude Models.
\newblock \url{https://docs.anthropic.com/en/docs/about-claude/models}.
\newblock Accessed: 2024-06-20.

\bibitem[{Austin et~al.(2021)Austin, Odena, Nye, Bosma, Michalewski, Dohan, Jiang, Cai, Terry, Le et~al.}]{austin2021program}
Austin, J.; Odena, A.; Nye, M.; Bosma, M.; Michalewski, H.; Dohan, D.; Jiang, E.; Cai, C.; Terry, M.; Le, Q.; et~al. 2021.
\newblock Program synthesis with large language models.
\newblock \emph{arXiv preprint arXiv:2108.07732}.

\bibitem[{Bai et~al.(2023)Bai, Zhao, Shi, Wei, Wu, and He}]{bai2023fairbench}
Bai, Y.; Zhao, J.; Shi, J.; Wei, T.; Wu, X.; and He, L. 2023.
\newblock FairBench: A Four-Stage Automatic Framework for Detecting Stereotypes and Biases in Large Language Models.
\newblock \emph{arXiv preprint arXiv:2308.10397}.

\bibitem[{Chen et~al.(2021)Chen, Tworek, Jun, Yuan, de~Oliveira~Pinto, Kaplan, Edwards, Burda, Joseph, Brockman, Ray, Puri, Krueger, Petrov, Khlaaf, Sastry, Mishkin, Chan, Gray, Ryder, Pavlov, Power, Kaiser, Bavarian, Winter, Tillet, Such, Cummings, Plappert, Chantzis, Barnes, Herbert-Voss, Guss, Nichol, Paino, Tezak, Tang, Babuschkin, Balaji, Jain, Saunders, Hesse, Carr, Leike, Achiam, Misra, Morikawa, Radford, Knight, Brundage, Murati, Mayer, Welinder, McGrew, Amodei, McCandlish, Sutskever, and Zaremba}]{chen2021evaluating}
Chen, M.; Tworek, J.; Jun, H.; Yuan, Q.; de~Oliveira~Pinto, H.~P.; Kaplan, J.; Edwards, H.; Burda, Y.; Joseph, N.; Brockman, G.; Ray, A.; Puri, R.; Krueger, G.; Petrov, M.; Khlaaf, H.; Sastry, G.; Mishkin, P.; Chan, B.; Gray, S.; Ryder, N.; Pavlov, M.; Power, A.; Kaiser, L.; Bavarian, M.; Winter, C.; Tillet, P.; Such, F.~P.; Cummings, D.; Plappert, M.; Chantzis, F.; Barnes, E.; Herbert-Voss, A.; Guss, W.~H.; Nichol, A.; Paino, A.; Tezak, N.; Tang, J.; Babuschkin, I.; Balaji, S.; Jain, S.; Saunders, W.; Hesse, C.; Carr, A.~N.; Leike, J.; Achiam, J.; Misra, V.; Morikawa, E.; Radford, A.; Knight, M.; Brundage, M.; Murati, M.; Mayer, K.; Welinder, P.; McGrew, B.; Amodei, D.; McCandlish, S.; Sutskever, I.; and Zaremba, W. 2021.
\newblock Evaluating Large Language Models Trained on Code.
\newblock arXiv:2107.03374.

\bibitem[{Chen, Cheung, and Yiu(2020)}]{chen2020metamorphic}
Chen, T.~Y.; Cheung, S.~C.; and Yiu, S.~M. 2020.
\newblock Metamorphic testing: a new approach for generating next test cases.
\newblock \emph{arXiv preprint arXiv:2002.12543}.

\bibitem[{Chen et~al.(2024)Chen, Zhang, Sarro, and Harman}]{fairness}
Chen, Z.; Zhang, J.~M.; Sarro, F.; and Harman, M. 2024.
\newblock Fairness Improvement with Multiple Protected Attributes: How Far Are We?
\newblock In \emph{Proceedings of the IEEE/ACM 46th International Conference on Software Engineering}, ICSE '24.

\bibitem[{Corbett-Davies et~al.(2017)Corbett-Davies, Pierson, Feller, Goel, and Huq}]{corbett2017algorithmic}
Corbett-Davies, S.; Pierson, E.; Feller, A.; Goel, S.; and Huq, A. 2017.
\newblock Algorithmic decision making and the cost of fairness.
\newblock In \emph{Proceedings of the 23rd acm sigkdd international conference on knowledge discovery and data mining}, 797--806.

\bibitem[{Dejanovi{\'c} et~al.(2017)Dejanovi{\'c}, Vaderna, Milosavljevi{\'c}, and Vukovi{\'c}}]{Dejanovic2017}
Dejanovi{\'c}, I.; Vaderna, R.; Milosavljevi{\'c}, G.; and Vukovi{\'c}, {\v{Z}}. 2017.
\newblock Textx: a python tool for domain-specific languages implementation.
\newblock \emph{Knowledge-based systems}, 115: 1--4.

\bibitem[{Dhamala et~al.(2021)Dhamala, Sun, Kumar, Krishna, Pruksachatkun, Chang, and Gupta}]{dhamala2021bold}
Dhamala, J.; Sun, T.; Kumar, V.; Krishna, S.; Pruksachatkun, Y.; Chang, K.-W.; and Gupta, R. 2021.
\newblock Bold: Dataset and metrics for measuring biases in open-ended language generation.
\newblock In \emph{Proceedings of the 2021 ACM conference on fairness, accountability, and transparency}, 862--872.

\bibitem[{D{\'\i}az et~al.(2018)D{\'\i}az, Johnson, Lazar, Piper, and Gergle}]{diaz2018addressing}
D{\'\i}az, M.; Johnson, I.; Lazar, A.; Piper, A.~M.; and Gergle, D. 2018.
\newblock Addressing age-related bias in sentiment analysis.
\newblock In \emph{Proceedings of the 2018 chi conference on human factors in computing systems}, 1--14.

\bibitem[{Galhotra, Brun, and Meliou(2017)}]{galhotra2017fairness}
Galhotra, S.; Brun, Y.; and Meliou, A. 2017.
\newblock Fairness testing: testing software for discrimination.
\newblock In \emph{Proceedings of the 2017 11th Joint meeting on foundations of software engineering}, 498--510.

\bibitem[{Gallegos et~al.(2023)Gallegos, Rossi, Barrow, Tanjim, Kim, Dernoncourt, Yu, Zhang, and Ahmed}]{gallegos2023bias}
Gallegos, I.~O.; Rossi, R.~A.; Barrow, J.; Tanjim, M.~M.; Kim, S.; Dernoncourt, F.; Yu, T.; Zhang, R.; and Ahmed, N.~K. 2023.
\newblock Bias and fairness in large language models: A survey.
\newblock \emph{arXiv preprint arXiv:2309.00770}.

\bibitem[{Google(2023)}]{codechat}
Google. 2023.
\newblock Code Chat.
\newblock \url{https://cloud.google.com/vertex-ai/generative-ai/docs/model-reference/code-chat}.
\newblock Accessed: 2024-06-20.

\bibitem[{Huang et~al.(2023)Huang, Bu, Zhang, Xie, Chen, and Cui}]{Huang2023BiasTA}
Huang, D.; Bu, Q.; Zhang, J.; Xie, X.; Chen, J.; and Cui, H. 2023.
\newblock Bias Testing and Mitigation in LLM-based Code Generation.
\newblock \emph{https://api.semanticscholar.org/CorpusID:262824773}.

\bibitem[{Li et~al.(2023)Li, Allal, Zi, Muennighoff, Kocetkov, Mou, Marone, Akiki, Li, Chim et~al.}]{li2023starcoder}
Li, R.; Allal, L.~B.; Zi, Y.; Muennighoff, N.; Kocetkov, D.; Mou, C.; Marone, M.; Akiki, C.; Li, J.; Chim, J.; et~al. 2023.
\newblock StarCoder: may the source be with you!
\newblock \emph{arXiv preprint arXiv:2305.06161}.

\bibitem[{Liang et~al.(2021)Liang, Wu, Morency, and Salakhutdinov}]{liang2021towards}
Liang, P.~P.; Wu, C.; Morency, L.-P.; and Salakhutdinov, R. 2021.
\newblock Towards understanding and mitigating social biases in language models.
\newblock In \emph{International Conference on Machine Learning}, 6565--6576. PMLR.

\bibitem[{Liu et~al.(2019)Liu, Dacon, Fan, Liu, Liu, and Tang}]{liu2019does}
Liu, H.; Dacon, J.; Fan, W.; Liu, H.; Liu, Z.; and Tang, J. 2019.
\newblock Does gender matter? towards fairness in dialogue systems.
\newblock \emph{arXiv preprint arXiv:1910.10486}.

\bibitem[{Liu et~al.(2023)Liu, Chen, Gao, Su, Zhang, Zan, Lou, Chen, and Ho}]{liu2024uncovering}
Liu, Y.; Chen, X.; Gao, Y.; Su, Z.; Zhang, F.; Zan, D.; Lou, J.-G.; Chen, P.-Y.; and Ho, T.-Y. 2023.
\newblock Uncovering and quantifying social biases in code generation.
\newblock \emph{Advances in Neural Information Processing Systems}, 36.

\bibitem[{Meade, Poole-Dayan, and Reddy(2021)}]{meade2021empirical}
Meade, N.; Poole-Dayan, E.; and Reddy, S. 2021.
\newblock An empirical survey of the effectiveness of debiasing techniques for pre-trained language models.
\newblock \emph{arXiv preprint arXiv:2110.08527}.

\bibitem[{M{\v{e}}chura(2022)}]{mechura-2022-taxonomy}
M{\v{e}}chura, M. 2022.
\newblock A taxonomy of bias-causing ambiguities in machine translation.
\newblock In \emph{Proceedings of the 4th Workshop on Gender Bias in Natural Language Processing (GeBNLP)}, 168--173.

\bibitem[{Meta(2024)}]{codellama}
Meta. 2024.
\newblock Code Llama 70B Instruct HF.
\newblock \url{https://huggingface.co/meta-llama/CodeLlama-70b-Instruct-hf}.
\newblock Accessed: 2024-06-20.

\bibitem[{Mozafari, Farahbakhsh, and Crespi(2020)}]{mozafari2020hate}
Mozafari, M.; Farahbakhsh, R.; and Crespi, N. 2020.
\newblock Hate speech detection and racial bias mitigation in social media based on BERT model.
\newblock \emph{PloS one}, 15(8): e0237861.

\bibitem[{Nadeem, Bethke, and Reddy(2020)}]{nadeem2020stereoset}
Nadeem, M.; Bethke, A.; and Reddy, S. 2020.
\newblock StereoSet: Measuring stereotypical bias in pretrained language models.
\newblock \emph{arXiv preprint arXiv:2004.09456}.

\bibitem[{Nangia et~al.(2020)Nangia, Vania, Bhalerao, and Bowman}]{nangia2020crows}
Nangia, N.; Vania, C.; Bhalerao, R.; and Bowman, S.~R. 2020.
\newblock CrowS-pairs: A challenge dataset for measuring social biases in masked language models.
\newblock \emph{arXiv preprint arXiv:2010.00133}.

\bibitem[{Nijkamp et~al.(2022)Nijkamp, Pang, Hayashi, Tu, Wang, Zhou, Savarese, and Xiong}]{nijkamp2022codegen}
Nijkamp, E.; Pang, B.; Hayashi, H.; Tu, L.; Wang, H.; Zhou, Y.; Savarese, S.; and Xiong, C. 2022.
\newblock Codegen: An open large language model for code with multi-turn program synthesis.
\newblock \emph{arXiv preprint arXiv:2203.13474}.

\bibitem[{OpenAI(2022)}]{openai}
OpenAI. 2022.
\newblock GPT-3.5 Turbo Models.
\newblock \url{https://platform.openai.com/docs/models/gpt-3-5-turbo}.
\newblock Accessed: 2024-06-20.

\bibitem[{Parrish et~al.(2021)Parrish, Chen, Nangia, Padmakumar, Phang, Thompson, Htut, and Bowman}]{parrish2021bbq}
Parrish, A.; Chen, A.; Nangia, N.; Padmakumar, V.; Phang, J.; Thompson, J.; Htut, P.~M.; and Bowman, S.~R. 2021.
\newblock BBQ: A hand-built bias benchmark for question answering.
\newblock \emph{arXiv preprint arXiv:2110.08193}.

\bibitem[{Rekabsaz and Schedl(2020)}]{rekabsaz2020neural}
Rekabsaz, N.; and Schedl, M. 2020.
\newblock Do neural ranking models intensify gender bias?
\newblock In \emph{Proceedings of the 43rd International ACM SIGIR Conference on Research and Development in Information Retrieval}, 2065--2068.

\bibitem[{Roziere et~al.(2023)Roziere, Gehring, Gloeckle, Sootla, Gat, Tan, Adi, Liu, Remez, Rapin et~al.}]{roziere2023code}
Roziere, B.; Gehring, J.; Gloeckle, F.; Sootla, S.; Gat, I.; Tan, X.~E.; Adi, Y.; Liu, J.; Remez, T.; Rapin, J.; et~al. 2023.
\newblock Code llama: Open foundation models for code.
\newblock \emph{arXiv preprint arXiv:2308.12950}.

\bibitem[{Sap et~al.(2019)Sap, Card, Gabriel, Choi, and Smith}]{sap-etal-2019-risk}
Sap, M.; Card, D.; Gabriel, S.; Choi, Y.; and Smith, N.~A. 2019.
\newblock The risk of racial bias in hate speech detection.
\newblock In \emph{Proceedings of the 57th annual meeting of the association for computational linguistics}, 1668--1678.

\bibitem[{Sheng et~al.(2020)Sheng, Chang, Natarajan, and Peng}]{sheng2020towards}
Sheng, E.; Chang, K.-W.; Natarajan, P.; and Peng, N. 2020.
\newblock Towards controllable biases in language generation.
\newblock \emph{arXiv preprint arXiv:2005.00268}.

\bibitem[{Steed et~al.(2022)Steed, Panda, Kobren, and Wick}]{steed-etal-2022-upstream}
Steed, R.; Panda, S.; Kobren, A.; and Wick, M. 2022.
\newblock Upstream mitigation is not all you need: Testing the bias transfer hypothesis in pre-trained language models.
\newblock In \emph{Proceedings of the 60th Annual Meeting of the Association for Computational Linguistics (Volume 1: Long Papers)}, 3524--3542.

\bibitem[{Wan et~al.(2023)Wan, Wang, He, Gu, Bai, and Lyu}]{wan2023biasasker}
Wan, Y.; Wang, W.; He, P.; Gu, J.; Bai, H.; and Lyu, M.~R. 2023.
\newblock Biasasker: Measuring the bias in conversational ai system.
\newblock In \emph{Proceedings of the 31st ACM Joint European Software Engineering Conference and Symposium on the Foundations of Software Engineering}, 515--527.

\bibitem[{Wikipedia(2024)}]{enwiki:1224247091}
Wikipedia. 2024.
\newblock Student's t-test.
\newblock \url{https://en.wikipedia.org/wiki/Student%27s_t-test}.
\newblock Accessed: 2024-06-20.

\bibitem[{Yang et~al.(2022)Yang, Yi, Li, Liu, and Xie}]{yang2022unified}
Yang, Z.; Yi, X.; Li, P.; Liu, Y.; and Xie, X. 2022.
\newblock Unified detoxifying and debiasing in language generation via inference-time adaptive optimization.
\newblock \emph{arXiv preprint arXiv:2210.04492}.

\bibitem[{Zhang et~al.(2023)Zhang, Sun, Wang, and Sun}]{zhang2023testsgd}
Zhang, M.; Sun, J.; Wang, J.; and Sun, B. 2023.
\newblock TESTSGD: Interpretable testing of neural networks against subtle group discrimination.
\newblock \emph{ACM Transactions on Software Engineering and Methodology}, 32(6): 1--24.

\bibitem[{Zhao et~al.(2023)Zhao, Fang, Pan, Yin, and Pechenizkiy}]{zhao2023gptbias}
Zhao, J.; Fang, M.; Pan, S.; Yin, W.; and Pechenizkiy, M. 2023.
\newblock GPTBIAS: A comprehensive framework for evaluating bias in large language models.
\newblock \emph{arXiv preprint arXiv:2312.06315}.

\end{thebibliography}






\end{document}